\documentclass[aps,pre,twocolumn,groupedaddress,showpacs]{revtex4}

\usepackage{graphicx} 

\newcommand{\odn}[3]{\frac{d^{#3}#1}{d#2^{#3}}}

\newcommand{\pdn}[3]{\frac{\partial^{#3}#1}{\partial#2^{#3}}}
\newcommand{\pd}[2]{\frac{\partial#1}{\partial#2}}

\usepackage{amsmath}
\newcommand{\sech}{\operatorname{sech}}

\newcommand{\Ai}{\operatorname{Ai}}
\newcommand{\Bi}{\operatorname{Bi}}
\begin{document}


\title{Nonlocal effects in high energy charged particle beams}


\author{Pontus Johannisson}
\email[]{pontus.johannisson@elmagn.chalmers.se}
\affiliation{Department of Electromagnetics, Chalmers University of
Technology, SE--412~96~G\"oteborg, Sweden}

\author{Dan Anderson}
\affiliation{Department of Electromagnetics, Chalmers University of
Technology, SE--412~96~G\"oteborg, Sweden}

\author{Mietek Lisak}
\affiliation{Department of Electromagnetics, Chalmers University of
Technology, SE--412~96~G\"oteborg, Sweden}

\author{Mattias Marklund}
\affiliation{Department of Electromagnetics, Chalmers University of
Technology, SE--412~96~G\"oteborg, Sweden}

\author{Renato Fedele}
\affiliation{Dipartimento di Fisica, Universita` Federico II and INFN,
Complesso Universitario di M.~S.~Angelo, Via Cintia, Naples, I--80126
Italy}

\author{Arkadi Kim}
\affiliation{Institute of Applied Physics, Russian Academy of
Sciences, 603950 Nizhny Novgorod, Russia}


\date{\today}

\begin{abstract}
Within the framework of the thermal wave model, an investigation is
made of the longitudinal dynamics of high energy charged particle
beams. The model includes the self-consistent interaction between the
beam and its surroundings in terms of a nonlinear coupling impedance,
and when resistive as well as reactive parts are included, the
evolution equation becomes a generalised nonlinear Schr\"odinger
equation including a nonlocal nonlinear term. The consequences of the
resistive part on the propagation of particle bunches are examined
using analytical as well as numerical methods.
\end{abstract}


\maketitle


\section{Introduction}
The thermal wave model (TWM), \cite{fedele, fedele_pla1, fedele_pla2},
describes the dynamics of high energy charged particle beams in
accelerators.  In the TWM approach, the beam is characterised by a
complex-valued wave function, which satisfies a Schr\"odinger-like
evolution equation, where the beam emittance replaces Planck's
constant, and the intensity of the wave function corresponds to the
beam particle density.  The Schr\"odinger potential, which describes
the interaction between the beam and its surroundings, can be
expressed in terms of a coupling impedance, and due to collective
effects, the coupling is a nonlinear function of the beam density. For
purely reactive impedances, the TWM equation reduces to the well-known
nonlinear Schr\"odinger equation. However, by including also the
resistive part, the evolution equation becomes a generalised
Schr\"odinger equation containing a new term, which is both nonlinear
and nonlocal. The modulational instability properties of this new
equation have been analysed previously, \cite{anderson_I}, and have
been shown to agree with results obtained using classical approaches,
including kinetic effects like Landau damping, \cite{fedele2}.

In the present work we consider the longitudinal dynamics of particle
bunches under the influence of the coupling impedance. Since the case
with a purely reactive impedance is well known, main emphasis is on
the situation where the resistive part is included.  The dynamical
evolution then proceeds as a competition between linear diffraction,
nonlinear self-focusing/defocusing, and nonlocal self-steepening.  It
is found that the bunch is accelerated/decelerated, and the
self-steepening effect makes the pulse shape asymmetric with an
extended tail, and eventually a wave-breaking phenomenon can appear on
the steepening edge.  An approximate solution for the dynamics of
soliton-shaped bunches in the presence of a small resistive impedance
is found using variational methods, and a perturbation solution for
the initial evolution of a particle bunch is given. This illustrates
the interplay between the different effects and is qualitatively in
agreement with results obtained by other means, as summarised, e.g.,
in \cite{ruth}.  No stationary solutions with finite particle number
are possible, but semi-infinite shocks and pulses with extended wake
fields are found.  The analytical predictions are confirmed by
numerical simulations of the full generalised nonlinear Schr\"odinger
equation.

\section{The generalised nonlinear Schr\"odinger equation}
Within the TWM, the longitudinal dynamics of particle bunches are
analysed in terms of a complex beam wave function $\psi(x, z)$, where
$z$ is the distance of propagation and $x$ is the longitudinal
extension of the particle beam, measured in the moving frame of
reference.  The particle density, $n(x, z)$, is related to the wave
function according to $n(x, z) = |\psi(x, z)|^2$, see
\cite{fedele}. The evolution of the beam is described by the
generalised Schr\"odinger equation
\begin{equation}
  i \pd{\psi}{z} = \alpha \pdn{\psi}{x}{2} + \kappa |\psi|^2 \psi + \mu
  \psi \int_{-\infty}^{x} |\psi(\xi, z)|^2 \, d\xi,
  \label{eq_nlse_nonlocal}
\end{equation}
where the longitudinal diffraction parameter, $\alpha$, can be both
positive and negative, depending on the phase slip parameter,
\cite{lawson}, and the nonlinear parameters, $\kappa$ and $\mu$, are
proportional to the imaginary (reactive) and real (resistive) parts,
respectively, of the coupling impedance.

In the case $\mu = 0$, Eq.~(\ref{eq_nlse_nonlocal}) reduces to the
fundamental nonlinear Schr\"odinger equation for which a wealth of
information is available. In particular, depending on the sign of
the product $\alpha \kappa$, the nonlinearity will either
counteract ($\alpha \kappa > 0$) or enhance ($\alpha \kappa < 0$)
the diffractive broadening. Furthermore, the velocity of the
particle bunch is left unchanged, and no asymmetry is introduced
on an initially symmetric bunch. Of special interest is the case
$\alpha \kappa > 0$, when shape-preserving soliton solutions are
possible as a balance between linear diffraction and nonlinear
self-focusing effects.

\begin{figure}
\includegraphics[width=0.98\linewidth]{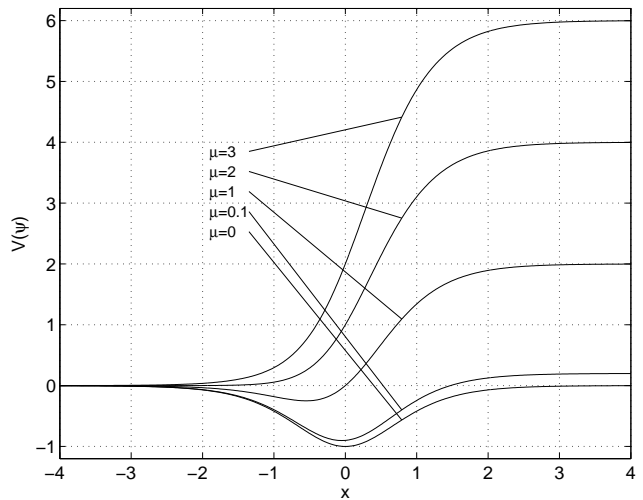}
\caption{\label{fig_potential_plot}A qualitative plot of the
potential, Eq.~(\ref{eq_nlse_potential}), for a pulse-shaped field
using different values of $\mu$.}
\end{figure}

However, the properties of the full Eq.~(\ref{eq_nlse_nonlocal}) are
not known and will be the subject of the present work. In order to see
the physical significance of the new term, it is instructive to
qualitatively discuss the nonlinear potential, $V(\psi)$, which in the
case $\alpha < 0$ is given by
\begin{equation}
  V(\psi) = \kappa |\psi|^2 + \mu \int_{-\infty}^{x} |\psi|^2 \, d\xi.
  \label{eq_nlse_potential}
\end{equation}
Consider first the case $\mu \to 0$ and $\alpha \kappa > 0$, i.e.,
assume that also $\kappa < 0$. Then, sech-shaped solutions form a
well-shaped potential that allows bound states, solitons, to
exist. The fundamental soliton solution corresponding to
Eq.~(\ref{eq_nlse_nonlocal}) is
\begin{equation}
  \psi = A_0 \sech (a x) e^{-i \delta z}, \quad a = \sqrt{\frac{\kappa
  A_0^2}{2 \alpha}}, \quad \delta = \frac{\kappa A_0^2}{2}.
  \label{eq_fundamental_soliton}
\end{equation}
The nonlocal part of the potential introduced by $\mu \neq 0$
contributes a monotonous term to the potential and creates an
asymmetry.  A qualitative plot of the total potential corresponding to
a field shaped as the fundamental soliton (choosing $A_0 = 1$, $a =
1$, and $\kappa = -1$), is shown in Fig.~\ref{fig_potential_plot},
using different values for $\mu$.
It is clear that if $\mu$ is small, the evolution of an initially
soliton-shaped pulse should involve an acceleration in a direction
determined by the sign of $\mu$, but the change of the pulse shape can
be expected to be slow due to the similarities with the conditions for
soliton propagation. For large $\mu$, however, it is obvious that
there can be no pulse-shaped stationary solutions, since the total
potential then is monotonous, and thus is unable to provide the
compression effects needed for confinement. By noticing that the slope
of the potential varies over the pulse, and that the more intense
parts are accelerated/decelerated more, strong internal pulse dynamics
is anticipated. Our subsequent analysis will confirm this intuitive
picture.

\section{Perturbed soliton dynamics}
From the potential picture it is expected that one of the main
effects of the nonlocal term is to induce an acceleration of an
initially stationary pulse. A more quantitative analysis of this
effect can be carried out by investigating the adiabatic evolution
of the soliton solution, Eq.~(\ref{eq_fundamental_soliton}), in
the presence of a small but finite value of $\mu$. This can
conveniently be done using a direct variational approach, see,
e.g., \cite{Pramana}. A suitable trial function is
\begin{equation}
  \psi_T(x, z) = A \sech[a(x - M)] e^{i[C(x - M) + D]},
\end{equation}
where $A(z)$, $a(z)$, $C(z)$, $D(z)$, and $M(z)$ are unknown parameter
functions to be determined by the variational procedure. We emphasise
that this ansatz function neglects any asymmetric pulse deformations
and consequently can model only part of the dynamical evolution. Using
Ritz optimisation, these parameter functions can be determined and the
following approximate solution is obtained
\begin{eqnarray}
  \psi_T &=& A_0 \sech{\left( \sqrt{\frac{\kappa A_0^2}{2 \alpha}} \xi
  \right)} \nonumber \\
  &&\times \exp \left\{ i \left[ -\frac{2 \mu A_0^2}{3} z \xi -
  \frac{4 \alpha \mu^2 A_0^4}{27} z^3 \right. \right. \nonumber \\
  &&\left. \left. \hspace{10.1mm} - \left(\mu A_0^2 \sqrt{\frac{2
  \alpha}{\kappa A_0^2}} + \frac{\kappa A_0^2}{2} \right) z \right]
  \right\},
  \label{eq_variational_result}
\end{eqnarray}
where
\begin{equation}
  \xi = x - \frac{2 \alpha \mu A _0^2}{3}z^2.
\label{eqgamma}
\end{equation}
This solution is consistent with the classical NLS equation in the
sense that the fundamental soliton,
Eq.~(\ref{eq_fundamental_soliton}), is recovered in the limit when
$\mu \to 0$.  The solution in the general case describes a soliton
being accelerated in the original frame of reference. The
acceleration, $\gamma$, is given by $\gamma = 4 \alpha \mu A_0^2/3$
and the concomitant shift of the group-velocity is associated with a
frequency shift proportional to $z$.

\begin{figure}
\includegraphics[width=0.98\linewidth]{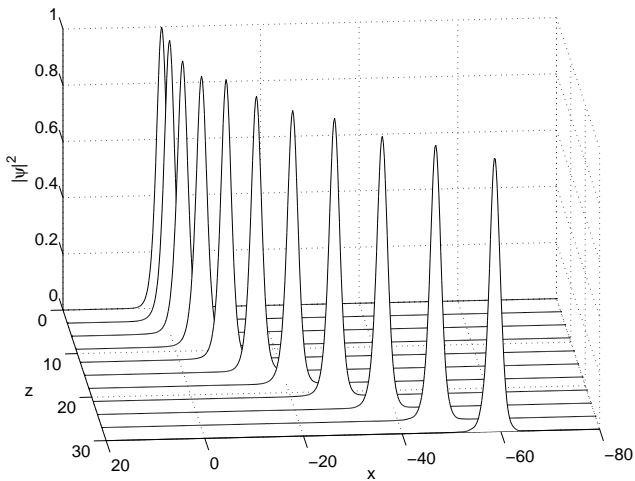}
\caption{\label{fig_mem_sim_L_30_mu_0.1}The numerically obtained
  dynamics of an initially soliton-shaped pulse using a weak nonlocal
  term.
}
\end{figure}

In order to check the approximate analytical solutions, but also to
obtain results in parameter ranges where analytical solutions are not
available, Eq.~(\ref{eq_nlse_nonlocal}) has been solved
numerically. For this purpose, the standard split-step Fourier method
for handling the NLSE is modified to include the effects of the
nonlocal term. An example of the dynamics caused by a weak nonlocal
term is seen in Fig.~\ref{fig_mem_sim_L_30_mu_0.1}.  The initial
pulse, which is centred around $x = 0$, is the fundamental soliton,
making a comparison with the variational result simple. The parameters
are $\alpha = -1$, $\kappa = -1$, $\mu = 0.1$, and $A_0 = 1$, and the
propagation distance is $L = 30$.  The pulse is accelerated towards
negative $x$ coordinates, but the shape is only weakly
distorted. Thus, for small values of $\mu$, the variational result,
which predicts the final centre position to be $x = -60$, describes
the propagation in an excellent way.
\begin{figure}
\includegraphics[width=0.98\linewidth]{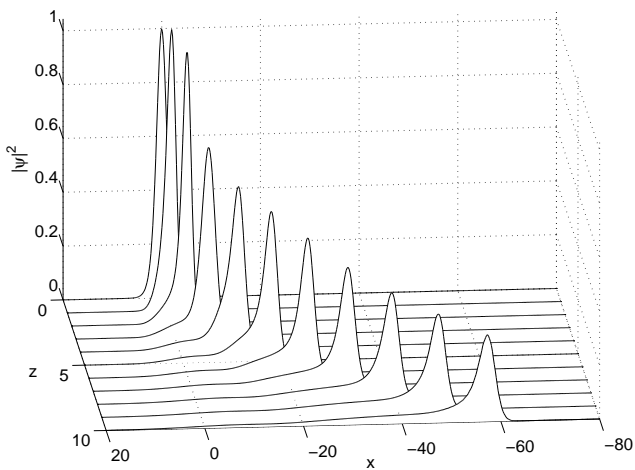}
\caption{\label{fig_mem_sim_L_10_mu_1}By increasing the strength of
the nonlocal term, strong pulse shape distortion is introduced.
}
\end{figure}

However, for larger values of $\mu$, the asymmetric deformation of the
pulse becomes more important and tends to leave an extended tail
behind the main pulse.  A numerical simulation result of this
situation is shown in Fig.~\ref{fig_mem_sim_L_10_mu_1}, where $\mu =
1$. The propagation distance is $L = 10$, making the final
$x$-position similar to the example above.  It is seen that the
internal dynamics of the pulse is significantly stronger; the peak
power decreases, and a wake field is developed. Again using the
variational result, the final position is expected to be $x = -66.7$,
but the decaying peak power makes the potential less steep during
propagation, which causes the acceleration to decrease.

Clearly, situations where $\mu$ is large can not be analysed by a
variational approach based on the soliton ansatz. In order to study
the deformation dynamics in some more analytical detail, we will
instead use a perturbation analysis.

\section{Perturbation analysis}
Although a general solution of Eq.~(\ref{eq_nlse_nonlocal}) based on
analytical methods is not possible, the initial dynamics can be
described using a perturbation analysis. For this purpose,
Eq.~(\ref{eq_nlse_nonlocal}) is rewritten as a coupled system in the
real amplitude, $A$, and the phase, $\theta$, of $\psi$ according to
\begin{eqnarray}
  \pd{A^2}{z} &=& 2 \alpha \pd{}{x} \left( A^2 \pd{\theta}{x} \right),
  \label{eq_da_dz} \\
  \pd{\theta}{z} &=& -\alpha \left[ \frac{1}{A} \pdn{A}{x}{2} - \left(
  \pd{\theta}{x} \right)^2 \right] - \kappa A^2 - \mu
  \int_{-\infty}^{x} \! \! \! \! A^2 \, d \xi. \ \ 
  \label{eq_dtheta_dz}
\end{eqnarray}
For an initially unchirped pulse, i.e., $\theta(x, 0) = 0$, the
initial amplitude modulation first creates a phase modulation
proportional to $z$, which then generates a subsequent change,
proportional to $z^2$, of the amplitude modulation. Let us consider
the case of the fundamental soliton,
Eq.~(\ref{eq_fundamental_soliton}), as initial field, since the
diffraction and the Kerr nonlinearity then balance each other. Thus,
as initial condition we consider $ A(x, z = 0) = A_0 \sech(a x)$,
where $a$ is related to $A_0$ according to
Eq.~(\ref{eq_fundamental_soliton}) and $\theta(x, z = 0) = 0$. Using
these in the right hand side of Eq.~(\ref{eq_dtheta_dz}), the initial
evolution of $\theta$ is obtained, and this solution can then be used
to find the lowest order modifications of $A$ from
Eq.~(\ref{eq_da_dz}).

However, as found using variational analysis, the pulse evolution is,
due to the effects of the nonlocal term, most conveniently described
in an accelerated coordinate system.  The proper value of the
acceleration can be taken from the previous section, but it is also
instructive to derive it using an analogy with Ehrenfest's theorem in
quantum mechanics. It is straightforward to show that the motion of
the mean position, $\langle x \rangle$, of the bunch obeys the
equation of motion
\begin{equation}
  \gamma_0 \equiv \odn{\langle x \rangle}{z}{2} = -2 \alpha \langle F
  \rangle,
\end{equation}
where the averaging is defined according to
\begin{equation}
  \langle f \rangle \equiv \frac{\int_{-\infty}^\infty f |\psi|^2 \,
  dx}{\int_{-\infty}^\infty |\psi|^2 \, dx},
\end{equation}
and the force is $F = -\partial V / \partial x$.  The acceleration
obtained in this way is identical to that derived using the
variational approach. Thus, $\gamma_0=\gamma$ and the new coordinate,
$\xi$, is defined according to Eq.~(\ref{eqgamma}). The amplitude and
phase are then obtained as
\begin{eqnarray}
  A &=& A_0 \sech (a \xi) \nonumber \\
  && \times \sqrt{1 + 4 \alpha a \mu A_0^2 \tanh (a \xi) \left[
  \sech^2 (a \xi) - \frac{1}{3} \right] z^2}, \ \ 
  \label{eq_pert_a_result} \\
  \theta &=& -\left\{ \frac{\kappa A_0^2}{2} + \frac{\mu A_0^2}{a}
  [\tanh(a \xi) + 1] \right\} z.
  \label{eq_pert_theta_result}
\end{eqnarray}
The first part of the phase, Eq.~(\ref{eq_pert_theta_result}), does
not depend on $\xi$, and is identical to the phase of the fundamental
soliton.  The second part is tanh-shaped, which is due to the form of
the nonlocal potential term.  As expected from the potential picture,
it is found that the amplitude becomes asymmetric.

\begin{figure}
\includegraphics[width=0.98\linewidth]{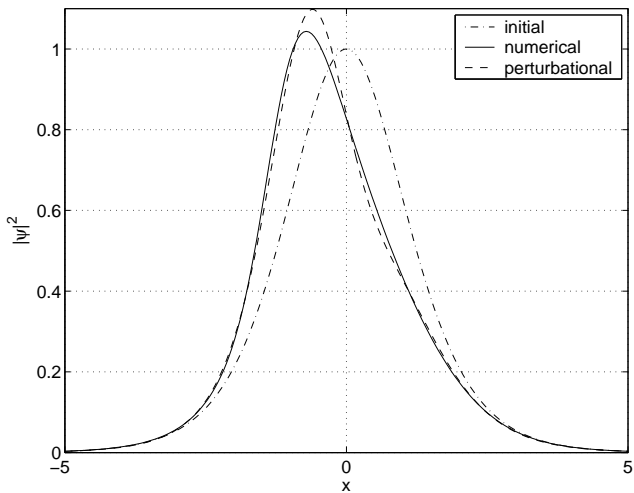}
\caption{\label{fig_mem_sim_L_0.2_mu_10}Comparison of the asymmetric
shapes predicted by analysis and numerical simulations, respectively.
}
\end{figure}

Due to the limited accuracy, the perturbation analysis can only be
applied within a certain propagation distance, obtained from
Eq.~(\ref{eq_pert_a_result}) as
\begin{equation}
  4 \alpha a \mu A_0^2 z^2 \ll 1 \quad \Rightarrow \quad
  z \ll 1/\sqrt{|4 \alpha a \mu A_0^2|}.
  \label{eq_pert_range}
\end{equation}
Using the same numerical parameters as above, no significant changes
in the amplitude are seen within that range. However, as seen in
Eq.~(\ref{eq_pert_range}), the application range decreases slowly as
$\mu$ increases. Thus, a large value, $\mu = 10$, has been used in
Fig.~\ref{fig_mem_sim_L_0.2_mu_10}, where the perturbation analysis is
compared with the numerically obtained result after a propagation
distance $L = 0.2$.  It is seen that the pulse is starting to ``lean
to the side'', and that the perturbation profile is in good agreement
with the numerical result, although its peak value is slightly too
large.

\section{Wave-breaking}
The nonlocal potential term, proportional to $\mu$ in
Eq.~(\ref{eq_nlse_potential}), gives rise to a force, $F(\xi) = -\mu
|\psi|^2$.  This implies that the central parts of the bunch are
affected by a stronger force than the wings, and will
accelerate/decelerate more. In fact, this is the basic mechanism
behind the steepening and the deformation of the bunch.  It is clear
that after a certain distance of propagation, the high amplitude parts
should overtake/be overtaken by the low amplitude parts of the
bunch. However, the finite diffraction will prohibit the development
of an infinite amplitude gradient, and the ``overtaking'' between
different parts of the bunch instead leads the appearance of
oscillations on the amplitude at the base of the steepening side of
the bunch. This feature is completely analogous to the wave-breaking
phenomenon in nonlinear defocusing Kerr media, \cite{tomlinson}, with
the difference that in the latter case, the corresponding force is an
odd function, which implies that the wave remains stationary and that
the wave-breaking phenomenon occurs symmetrically on both sides of the
pulse.

In order to estimate the length-scale of the wave-breaking phenomenon,
the perturbation solution for the amplitude,
Eq.~(\ref{eq_pert_a_result}), can be used.  Thus, the order of
magnitude of the wave-breaking distance, $z_{wb}$, is estimated as the
shortest propagation length for which the amplitude has a zero. This
is easily shown to occur at
$z_{wb} = \sqrt{3 / (4 |\alpha| a \mu A_0^2)}$.
It is interesting to note that this approach gives the same result as
the one used in \cite{anderson_wb}, which was based on the local
velocity shear in the pulse created by the nonlinearly induced chirp,
provided the latter is generalised to include the mean acceleration of
the pulse.

\begin{figure}
\includegraphics[width=0.98\linewidth]{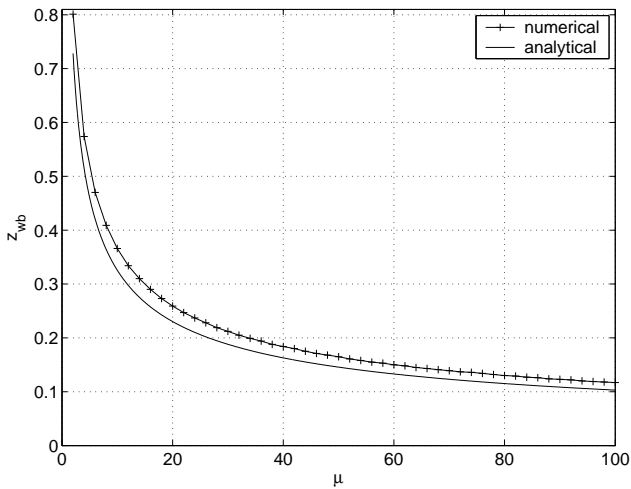}
\caption{\label{fig_wavebreak_distance2}The wave-breaking distance as
  predicted by analysis and numerical simulations, respectively.}
\end{figure}

By increasing the propagation distance in
Fig.~\ref{fig_mem_sim_L_0.2_mu_10}, oscillations on the amplitude will
start to occur at the base of the pulse on the steepening side, i.e.,
close to $x = -3$. Numerically we define the wave-breaking distance as
the propagation distance where the amplitude acquires a second
maximum. In Fig.~\ref{fig_wavebreak_distance2}, the analytical
prediction for the wave-breaking distance is compared with the result
of the numerical computations. The results show very good agreement,
although the numerical results tend to be somewhat larger than
predicted. However, since only an order-of-magnitude estimate has been
made, the result is quite satisfactory.  In particular, the analytic
result predicts very well how the wave-breaking distance scales with
$\mu$.

\section{Stationary solutions}
As already discussed, the purely reactive case, corresponding to $\mu
= 0$, allows a soliton solution, Eq.~(\ref{eq_fundamental_soliton}),
containing a finite number of particles. In order to investigate
whether similar stationary solutions exist also in the general case,
we return to Eqs.~(\ref{eq_da_dz}) and (\ref{eq_dtheta_dz}), which
describe the evolution of the (real) amplitude and the phase of the
wave, respectively. Based on our previous results, we will look for
solutions that are stationary in an accelerated frame of reference,
i.e., we introduce $\xi = x - \gamma z^2/2$, where the acceleration,
$\gamma$, now is unknown and has the character of an
eigenvalue. Stationarity implies that the amplitude depends only on
the coordinate $\xi$, i.e., $A = A(\xi)$.  The phase variation can
then be found explicitly, and the system becomes
\begin{eqnarray}
  &&\theta = \theta_0 + C_1 z - \frac{\gamma z \xi}{2 \alpha} -
  \frac{\gamma^2 z^3}{12 \alpha}, \\
  &&\alpha \odn{A}{\xi}{2} + C_1 A - \frac{\gamma \xi A}{2 \alpha} +
  \kappa A^3 + \mu A \int_{-\infty}^{\xi} \! \! \! \! A^2 \, d \xi' =
  0, \ \ 
  \label{eq_a_eqn_unnorm}
\end{eqnarray}
where $C_1$ is a constant, which acts as a second eigenvalue.
However, by rescaling $\alpha$, $\gamma$, $\kappa$, and $\mu$, the
equation can be rewritten as
\begin{equation}
  \alpha \odn{A}{\xi}{2} + A - \frac{\gamma \xi A}{2 \alpha} + \kappa
  A^3 + \mu A \int_{-\infty}^{\xi} \! \! \! \! A^2 \, d \xi' = 0.
  \label{eq_a_eqn_norm}
\end{equation}
In effect, this normalisation sets $C_1 = 1$, and the physical
significance of this can be found by letting $\mu \to 0$ (implying
also that $\gamma \to 0$).  The fundamental soliton is then recovered
from Eq.~(\ref{eq_a_eqn_unnorm}), and the phase is given by $C_1 z$.
Thus, setting $C_1 = 1$ corresponds to normalising with respect to the
propagation constant.

Assume that a pulse-shaped stationary solution exists.  This implies
that $A \to 0$ as $\xi \to -\infty$, and asymptotically the field
should satisfy the equation
\begin{equation}
\alpha \odn{A}{\xi}{2} + A - \frac{\gamma \xi A}{2 \alpha} = 0,
\end{equation}
which can be solved in terms of the Airy functions, $\Ai(x)$ and
$\Bi(x)$, as
\begin{equation}
A = D_1 \Ai \left( \frac{\gamma \xi - 2 \alpha}{\sqrt[3]{2 \alpha^2
\gamma^2}} \right) + D_2 \Bi \left( \frac{\gamma \xi - 2
\alpha}{\sqrt[3]{2 \alpha^2 \gamma^2}} \right).
\end{equation}
In order to obtain a solution containing a finite number of particles,
it is necessary that $\gamma <0$ and that $D_2=0$.  For pulse-like
solutions, the amplitude of the solution must also vanish as
$\xi\rightarrow\infty$ and the corresponding asymptotic equation is
\begin{equation}
\alpha \odn{A}{\xi}{2} + (1 + \mu W) A - \frac{\gamma \xi A}{2 \alpha}
= 0.
\end{equation}
Here, $W$ is the total number of particles, which has been assumed to
be finite, and the corresponding solution is
\begin{eqnarray}
A &=& D_3 \Ai \left( \frac{\gamma \xi - 2 (1 + \mu W)
\alpha}{\sqrt[3]{2 \alpha^2 \gamma^2}} \right)\\
&& + D_4 \Bi \left( \frac{\gamma \xi - 2 (1 + \mu W)
\alpha}{\sqrt[3]{2 \alpha^2 \gamma^2}} \right).
\end{eqnarray}
Since the acceleration has already been chosen to be negative, this
implies that the asymptotic solution will be the sum of two
oscillating Airy functions as $\xi \to \infty$. However, the total
number of particles of such a solution is infinite, and a
contradiction has been reached.  Thus, we conclude that there are no
stationary solutions to Eq.~(\ref{eq_nlse_nonlocal}) containing a
finite number of particles.

\begin{figure}
\includegraphics[width=0.98\linewidth]{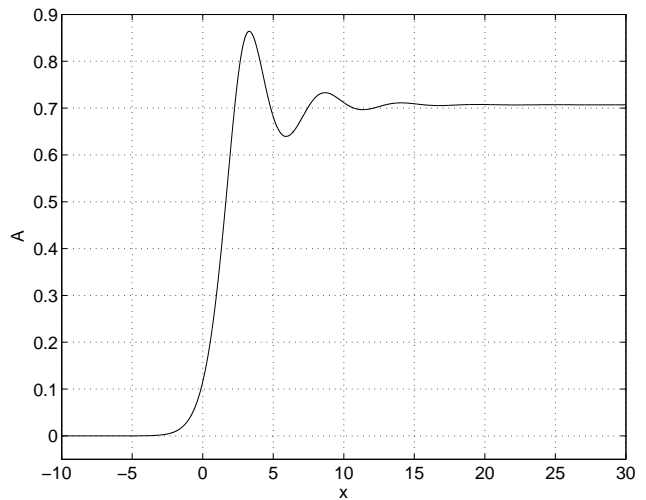}
\caption{\label{fig_step_func}A numerically obtained step-like
solution.}
\end{figure}

On the other hand, if the condition $A \to 0$ as $\xi \to \infty$ is
relaxed, step-like solutions can be found.  By assuming in
Eq.~(\ref{eq_a_eqn_norm}) that $A \to A_\infty$ when $\xi \to \infty$,
the integral term is asymptotically equal to $\mu A_\infty^3 \xi$.
This term can cancel the term that gives rise to the Airy solutions,
provided that $A_\infty = \sqrt{\gamma / (2 \alpha \mu)}$.
A solution of this type has been calculated numerically by using
$\alpha = -1$, $\kappa = -1$, $\gamma = -1$, and $\mu = 1$, and by
choosing the amplitude for the asymptotic solution for negative
$\xi$. The result has been plotted in Fig.~\ref{fig_step_func}, and it
is seen that the predicted value for the asymptotic amplitude,
$A_\infty = 1/\sqrt{2}$, is correct.

\section{Conclusion}

In conclusion, a generalised NLSE, describing the nonlinear
longitudinal dynamics of high energy charged particle beams in
accelerators within the TWM approach, has been analysed using both
analytical and numerical methods. It has been discussed in qualitative
physical terms how the inclusion of the resistive part of the coupling
impedance gives rise to both an acceleration and a deformation of the
particle bunch. These effects have been analysed analytically using
both a variational analysis and a direct perturbation analysis of the
initial dynamics, and the results have been shown to be in good
agreement with numerical simulations. It has also been shown that for
impedances with a large resistive part, the deformation leads to
self-steepening and eventually a wave-breaking phenomenon similar to
that occurring in nonlinear optics. The scale length for this effect
has been estimated and has also been shown to be in good agreement
with numerical results. Finally, it has been shown that no stationary
pulse-like solutions with a finite number of particles exist for the
generalised NLS equation, but semi-infinite shock solutions are
possible.


%


\bibliography{johannisson}

\end{document}